\documentclass[useAMS,usenatbib]{mn2e}
\usepackage{graphicx,amssymb,amsmath,times}
\title{Nearby supernova remnants and the cosmic-ray spectral hardening at high energies}
\author[Satyendra Thoudam]{Satyendra Thoudam\thanks{E-mail: s.thoudam@astro.ru.nl} and J\"{o}rg R. H\"{o}randel\\
Department of Astrophysics, IMAPP, Radboud University Nijmegen, P.O. Box 9010, 6500 GL Nijmegen, The Netherlands}

\begin{document}
\date{}
\pagerange{}
\maketitle
\label{firstpage}
\begin{abstract}
Recent measurements of cosmic-ray spectra of several individual nuclear species by the CREAM, TRACER, and ATIC experiments indicate a change in the spectral index of the power laws at TeV energies. Possible explanations among others include non linear diffusive shock acceleration of cosmic-rays, different cosmic-ray propagation properties at higher and lower energies in the Galaxy and the presence of nearby sources. In this paper, we show that if supernova remnants are the main sources of cosmic rays in our Galaxy, the effect of the nearby remnants can be responsible for the observed spectral changes. Using a rigidity dependent escape of cosmic-rays from the supernova remnants, we explain the apparent observed property that the hardening of the helium spectrum occurs at relatively lower energies as compared to the protons and also that the spectral hardening does not persist beyond $\sim (20-30)$ TeV energies.
\end{abstract}

\section{Introduction}
Recently, cosmic-ray (CR) measurements by the new-generation balloon-borne experiments such as the ATIC (Panov et al. 2007), CREAM (Yoon et al. 2011), and TRACER (Ave et al. 2008) seem to indicate that the CR spectrum deviates from a single power law. The spectra of all individual elements seem to be harder at TeV energies than at lower energies. Such a hardening is not easy to explain under the standard models of CR acceleration and their propagation in the Galaxy. Under the standard theory,  CRs below the knee ($\sim 3$ PeV) are considered to be produced by supernova remnant (SNR) shock waves by diffusive shock acceleration  mechanism (Bell 1978, Blandford $\&$ Eichler 1987). Such a mechanism naturally predicts a power law spectrum of $E^{-\gamma}$ with the index $\gamma=2$ for strong shocks. On the other hand, CR propagation in the Galaxy is considered to be of diffusive nature which is due to scattering by magnetic field irregularities and the CR self excited Alfven and hydromagnetic waves present in the Galaxy. Measurements of CR secondary-to-primary ratios indicate that the diffusion is energy dependent with the diffusion coefficient $D(E)\propto E^a$ with $a\approx (0.3-0.7)$. Under these considerations, the CR spectrum in the Galaxy is expected to follow a single power law with index $(\gamma+a)$ up to the knee, which do not seem to agree quite easily with the observed hardening at TeV energies.

The observed data can be explained if either the source spectrum or the diffusion index flattens at higher energies. Non linear diffusive shock acceleration theories where CRs modify the shock structure predict concave spectra (flatter at higher energies) at the shocks. But, the total spectrum injected into the interstellar medium which is the sum of the instantaneous spectra over the SNR life time is very close to a pure power-law (Caprioli et al. 2010). The concave signature can be even more diluted when summed over an ensemble of SNRs (Ptuskin $\&$ Zirakashvili 2005). From the propagation point of view, there are models which assume a harder or constant CR diffusion coefficient at higher energies in the Galaxy (Ave et al. 2009). Such models are motivated not only by the apparent flattening of the observed boron to carbon ratio above $\sim 100$ GeV energies, but also by the observed CR anisotropy which is almost independent of energy. Recently, it has also been proposed that dispersion in the spectral indices of CR source spectrum from many sources can also be responsible for the observed spectral hardening (Yuan et al. 2011).

Another possible explanation, as also pointed out in Ahn et al. 2010, is the presence of nearby sources. Erlykin $\&$ Wolfendale 2011 suggested that an extra component of CRs with a steep spectrum could be contributing below $\sim 200$ GeV/n while above that, the spectrum is entirely determined by a harder CR background. They proposed that the sources of the extra component could be in OB associations in the Local Bubble. Recently, Ohira $\&$ Ioka 2011 proposed that the hardening could be due to decreasing Mach number in hot superbubbles with multiple supernovae. In another recent work, Vladimirov et al. 2011 investigated several possible interpretations (including local source effect) for the observed  spectral features at low and high energies using the GALPROP propagation code. They also presented the possible effects on other observed properties such as CR anisotropy, isotopic ratios and the Galactic diffuse $\gamma$-ray emissions.  

In our present study, we investigate whether the spectral hardening observed at TeV energies could be an effect of the nearby SNRs. Although there has not been any direct detection of CRs from any sources, SNRs remain the most favorable candidates both theoretically as well as observationally. At least the presence of high energy particles up to few TeVs inside SNRs have been confirmed by the detections of non-thermal X-rays (Parizot et al. 2006) and TeV $\gamma$-rays from several SNRs (Aharonian et al. 2006, 2008a). Moreover, the detection of TeV electrons by the HESS experiment (Aharonian et al. 2008b) indicates the presence of one or more CR sources within a distance of $\sim 1$ kpc from us. If these sources produce both, electrons as well as nuclei, we expect to see some effects on the spectra of CR nuclei observed at the Earth.

\section {Model}
The diffusive propagation of CRs in the Galaxy neglecting the effects due to nuclear spallation can be described by the following equation,
\begin{equation}
\nabla\cdot(D\nabla N)+Q=\frac{\partial N}{\partial t}
\end{equation} 
where $N(\textbf{r},E,t)$ is the differential number density at a distance $\textbf{r}$ from the source at time $t$, $E$ is the kinetic energy/nucleon and $Q(\textbf{r},E,t)$ is the source term. The diffusion coefficient is taken as $D(\Re)=D_0(\Re/\Re_0)^a$ for $\Re>\Re_0$, where $\Re$ denotes the particle rigidity which is given by $\Re=AE/Z$ for charge $Z$ and mass number A. For our study, we consider two sets of values for $(D_0, \Re_0, a)$: one based on purely diffusion model (hereafter Model A) and the other based on models including CR reacceleration due to interstellar turbulence (hereafter Model B). We choose $(D_0, \Re_0, a)=(2.9, 3, 0.6)$ for Model A (Thoudam 2008) and $(5.8, 4, 0.33)$ for Model B (Strong et al. 2010), where $D_0$ is in units of $10^{28}$ cm$^2$ s$^{-1}$ and $\Re_0$ is in GV.

Under diffusive shock acceleration theory, CRs are confined within the remnant due to the magnetic turbulence generated by the CRs themselves. They can escape when their upstream diffusion length defined as $l_{diff} = D_s(E)/u_s$ is greater than the escape length from the shock front which is usually taken as $l_{esc}\approx 0.1 R_s$, where $u_s$ and $R_s$ denote the shock velocity and the shock radius respectively. In the Bohm diffusion limit, the upstream diffusion coefficient scales linearly with energy as $D_s(E)\propto E$ which implies that higher energy particles can escape the remnant at early times followed later by the lower energy ones. But, the exact energy dependence of $D_s$ is still not well understood and depends on some poorly known quantities which include the spectral distribution of the CR self-excited turbulence waves, the level of magnetic field amplification by the CRs and the dynamical reaction of CRs on the shock structure. Therefore, we follow a simple but reasonable parameterization for the CR escape time similar to that adopted by Gabici et al. 2009 as given below,
\begin{equation}
t_{esc}(\Re)=t_{sed}\left(\frac{\Re}{\Re_{max}}\right)^{-1/\alpha}
\end{equation}
where $t_{sed}$ denotes the start of the Sedov phase, $\Re_{max}$ denotes the maximum CR rigidity and $\alpha$ is a positive constant. We assume that the maximum CR energy accelerate by an SNR scales with the charge number $Z$ as $ZU_{max}$, where $U_{max}$ denotes the maximum kinetic energy of the protons which is taken as $1$ PeV for our study (Berezhko 1996). This scaling gives $\Re_{max}=1$ PV. In units of energy/nucleon, the maximum energy for helium is $E_{max}=0.5$ PeV/n.

Eq. (2) assumes that the highest energy CRs of all the species start escaping at the onset of the Sedov phase itself. Writing Eq. (2) in terms of total kinetic energy, it is easy to check that for the same kinetic energy, the escape time of CRs scales with the charge number as $Z^{1/\alpha}$, i.e, higher charged particles escape at relatively later stages of the SNR evolution. Thus, our escape model takes into account the general understanding of diffusive shock acceleration theory that higher charged particles can be confined for relatively longer duration within  the remnant. In terms of energy/nucleon, we can write Eq. (2) as
\begin{equation}
t_{esc}(E)=t_{sed}\left(\frac{AE}{Z\Re_{max}}\right)^{-1/\alpha}
\end{equation}
Eq. (3) shows that for the same energy/nucleon, all nuclei with charge $Z>1$ escape earlier than the protons by a factor of $(A/Z)^{-1/\alpha}$. We further assume that no particles remain confined after the shock completely dies out which we assume to occur when the SNR age $10^5$ yr. Taking this into account, the CR escape time for our study is taken as $T_{esc}(E)= \mathrm{min}\left[t_{esc}(E), 10^5 \mathrm{yr}\right]$. For detailed studies on particle escape from SNRs, see e.g., Ptuskin $\&$ Zirakashvili 2005, Caprioli et al. 2009 and Ohira et al. 2010.

The corresponding escape radius of CRs is calculated using the age-radius Sedov relation for SNRs as given below,
\begin{equation}
R_{esc}(E)=2.5u_0\;t_{sed}\left[\left(\frac{T_{esc}}{t_{sed}}\right)^{0.4}-0.6\right]
\end{equation}
where $u_0$ represents the initial shock velocity, i.e the velocity at $t=t_{sed}$.

The source term in Eq. (1) is taken as,
\begin{equation}
Q(\textbf{r},E,t)=\frac{q(E)}{A_{esc}}\delta(t-T_{esc})\delta(r-R_{esc})
\end{equation}
where $A_{esc}=4\pi R_{esc}^2$ denote the surface area of the SNR at the time when CRs of energy $E$ escape the remnant. It should be noted that our consideration of the rigidity dependent escape time and the finite source size are different from the commonly adopted burst-like point source approximation where CRs of all rigidities are assumed to escape at the same time from a point source. For CR study near the sources, the point source approximation can break down and it looks more realistic to take their sizes into account (Thoudam $\&$ H\"orandel 2011). Recently, such importance has also been highlighted in Ohira et al. 2011 in the study of $\gamma$-ray emission from SNRs interacting with molecular clouds.  

The source spectrum in Eq. (5) is taken as $q(E)=Aq(U)$ with $q(U)$ given by,
\begin{equation}
q(U)=k(U^2+2Um)^{-(\gamma+1)/2}(U+m)
\end{equation}
where $U=AE$ represents the particle total kinetic energy, $m$ is the rest mass energy and $k$ is the normalization constant which is related to the CR injection efficiency.

Solving Eq. (1), the spectrum at a distance $r_s$ from the SNR follows,
\begin{eqnarray}
N(r_s,E,t)=\frac{q(E)\,R_{esc}}{r_sA_{esc}\sqrt{\pi D(t-T_{esc})}}\mathrm{exp}\left[-\frac{\left(R_{esc}^2+r_s^2\right)}{4D(t-T_{esc})}\right]\nonumber\\
\times\;\mathrm{sinh}\left(\frac{r_sR_{esc}}{2D(t-T_{esc})}\right)\;
\end{eqnarray}
For high energy particles for which the diffusion radius defined as $r_{diff}=\sqrt{D(t-T_{esc})}$ is much larger than $(r_s, R_{esc})$, Eq. (7) follows a power-law of the form $N(E)\propto E^{-\left(\Gamma+\frac{3}{2}a\right)}$.

Eq. (7) can be used to calculate the CR spectra from the nearby SNRs. We choose proton and helium for our study and consider only those SNRs with distances $<1$ kpc from the Earth and ages $< 2\times 10^5$ yr. From the available literature, we found 10 SNRs listed as follows with their distances (kpc) and ages (yr)  respectively given in parentheses: Cygnus Loop $(0.54, 10^4)$, HB21 $(0.8, 1.9\times 10^4)$, HB9 $(0.8, 6.6\times 10^3)$, S147 $(0.8, 4.6\times 10^3)$, Vela $(0.3, 1.1\times 10^4)$, G299.2-2.9 $(0.5, 5\times 10^3)$, SN185 $(0.95, 1.8\times 10^3)$, Monogem $(0.3, 1.1\times 10^5)$, G114.3+0.3 $(0.7, 4.1\times 10^4)$ and Vela Junior $(0.75, 3.5\times 10^3)$.

In addition to the contributions from the nearby SNRs, we assume that there exists a steady CR background in the Galaxy which dominates the overall CR spectrum. For the CRs observed at the Earth, we assume that this background component consists of contributions from distant SNRs plus any other possible sources in the Galaxy. For our study, we obtain the background by fitting the observed CR spectrum between $(20-200)$ GeV/n. This is the energy region where the contamination from the nearby sources is expected to be less and at the same time, not much affected by the Solar modulation. In fact, it has been shown in Thoudam 2008 that the presence of nearby sources can produce stronger density fluctuations at higher energies than at lower energies because of the energy dependent nature of CR diffusion. Therefore, we believe that it is reasonable to assume that the low energy CRs that we observe at the Earth are not much affected by the presence of nearby SNRs and they largely represent the averaged background spectrum in the Galaxy. We will show in the following that this is indeed the most likely case.

\section{Results}
From the fit, the spectral indices of the background CRs are found to be $2.75\pm0.01$ for the protons and $2.68\pm0.02$ for the helium. The reason for the flatter helium spectrum is not properly understood. Recently, Blasi $\&$ Amato 2011 showed that the flatter helium spectrum with respect to the protons above $1$ TeV could be due to spallation effects. Later, Vladimirov et al. 2011 showed that such effects can lead to boron-to-carbon ratios and the anti-proton fluxes which are inconsistent with the observed data. Another possibility for the different spectral indices could be that the intrinsic source spectra itself are different. It could be due to different acceleration sites of protons and helium (Biermann et al. 2010) or inhomogeneous abundance of elements in superbubbles (Ohira $\&$ Ioka 2011). For our present study, we assume that CRs are injected into the Galaxy with different source indices. The index $\gamma$ for an individual species is chosen such that $(\gamma+a)$ is equal to the spectral index of the background obtain from the fit.

Before illustrating our results, we briefly discuss the choice of other model parameters involved in our calculations. Typically, $t_{sed}$ has values between $\sim (100- 10^3)$ yr depending on the gas density of the interstellar medium, mass of the ejecta and the energy output of the supernova explosion. For our study, we take $t_{sed}=500$ yr. We assume the initial shock velocity $u_0$ to be $10^9$ cm/s. This gives CR escape times from the SNRs in the range of $t_{esc}=(500-10^5)$ yr and the corresponding escape radii as $R_{esc}\sim(5-100)$ pc. Finally, we treat the escape parameter $\alpha$ and the injection efficiency of the protons (helium) hereafter denoted by $\epsilon_{p(he)}$ as free parameters. For our calculations, we will assume that all the parameters mentioned above are  same for all the SNRs.

\begin{figure}
\centering
\includegraphics*[width=0.315\textwidth,angle=-90,clip]{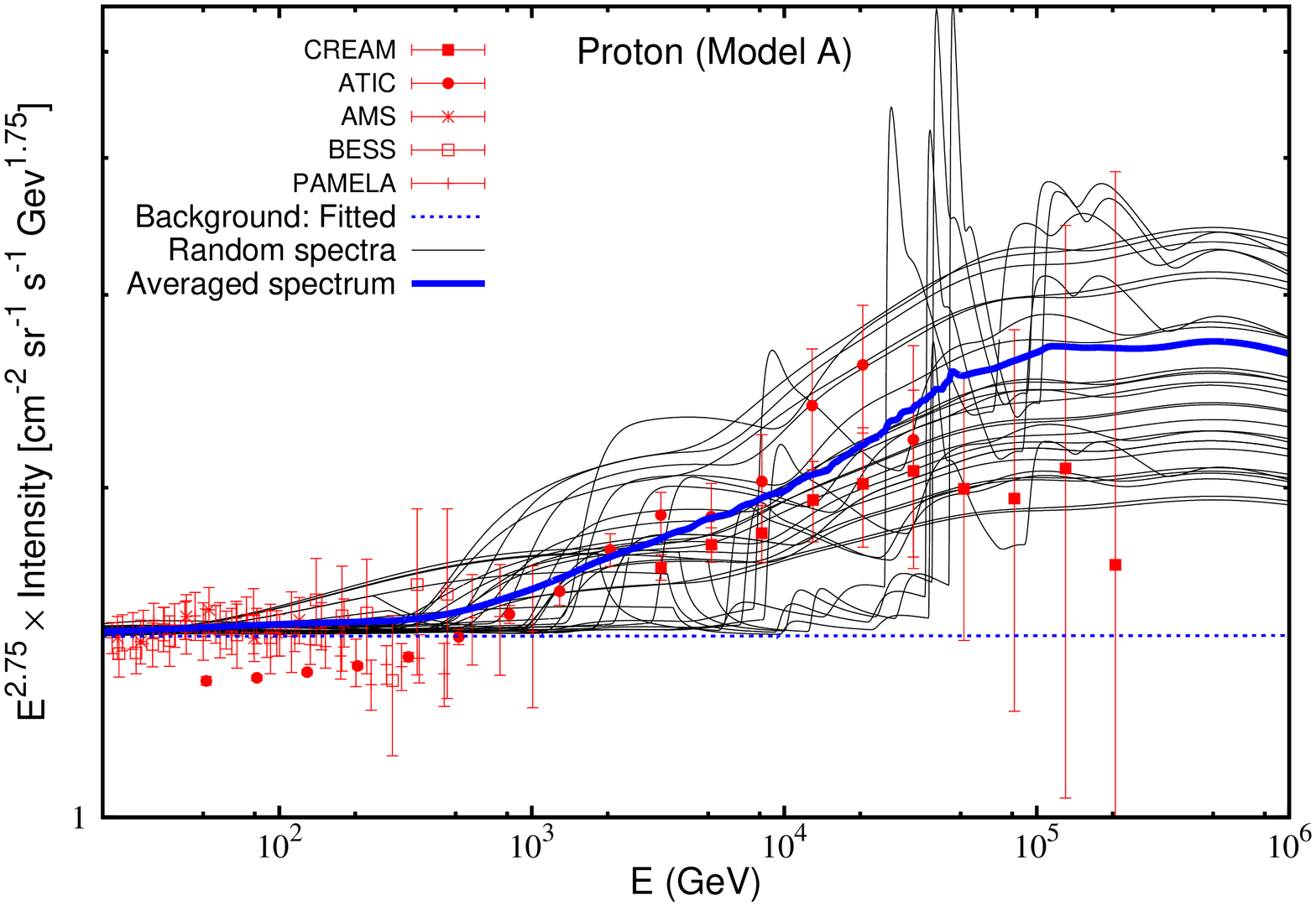}\\
\includegraphics*[width=0.315\textwidth,angle=-90,clip]{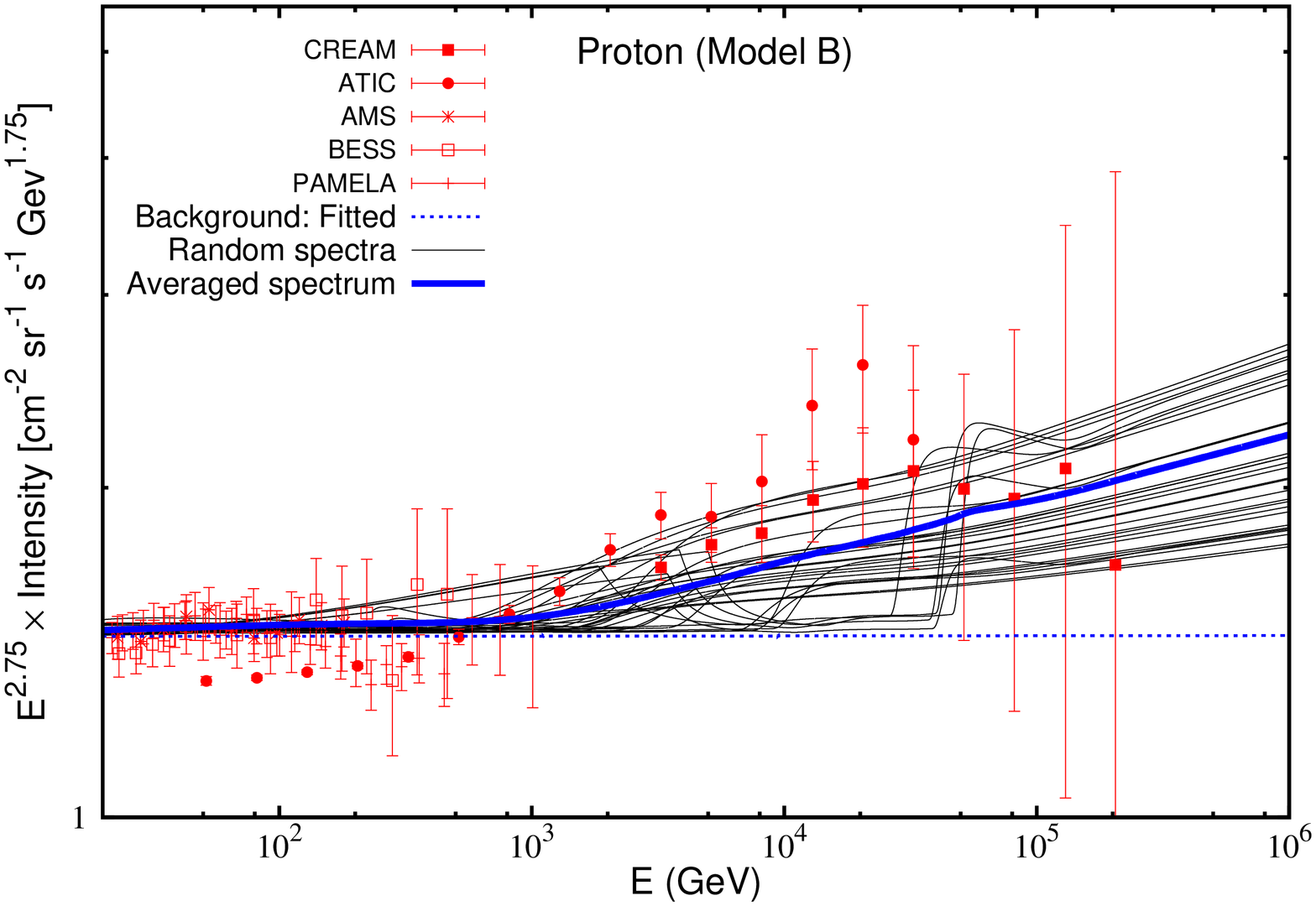}
\caption{\label {fig1} Proton spectra $(\times E^{2.75})$ for Model A (top panel) and Model B (bottom panel). The blue dotted line represents the background spectrum. The thin black lines represent an example of 30 random spectra calculated with proton escape parameters and the injection efficiencies in the range of $\alpha=(1-3)$ and $\epsilon_p=(5-25)\times 10^{49}$ ergs respectively. Each spectra is the sum of the background and the contribution from the nearby SNRs. The blue solid line represents the average of 200 random spectra. See text for data references and other details.}
\end{figure}

\begin{figure}
\centering
\includegraphics*[width=0.315\textwidth,angle=-90,clip]{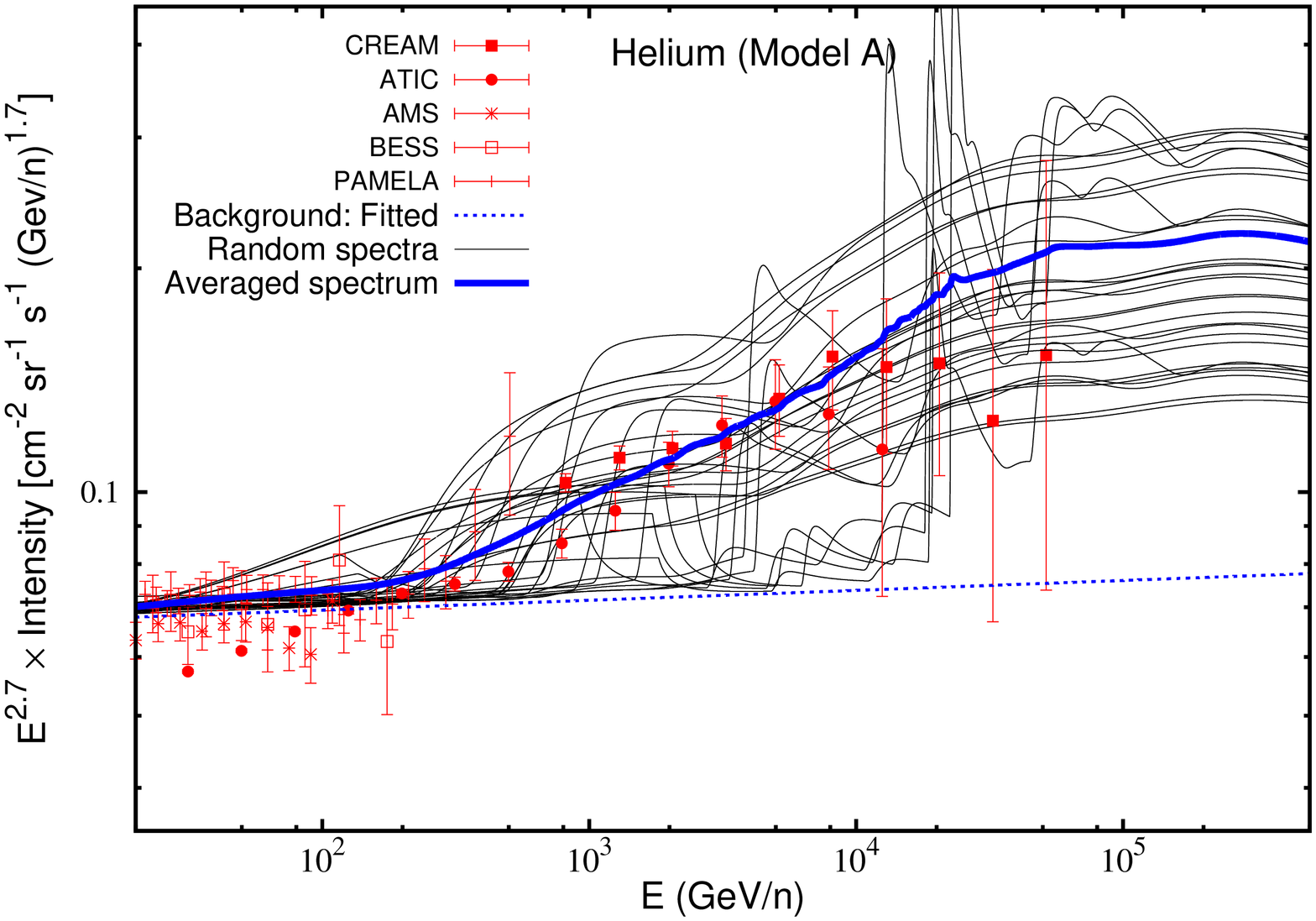}\\
\includegraphics*[width=0.315\textwidth,angle=-90,clip]{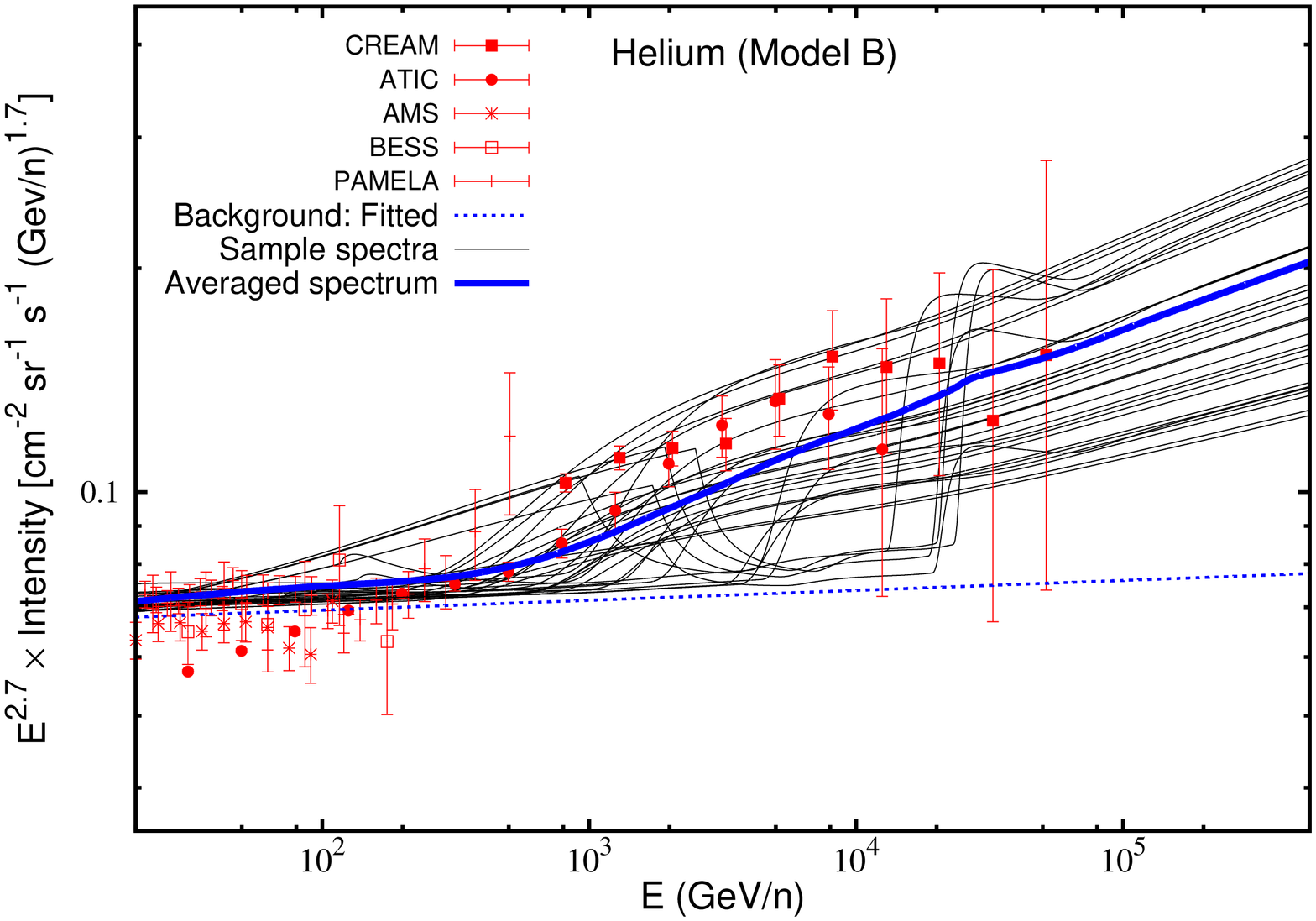}
\caption{\label {fig1} Helium spectra $(\times E^{2.7})$ for Model A (top panel) and Model B (bottom panel). The calculation assumes  injection efficiencies in the range of $\epsilon_{he}=(1-5)\times 10^{49}$ ergs. All other model parameters, results representation and the data references remain the same as in Fig.1.}
\end{figure}

Because of lack of precise informations on the values of $\alpha$ and $\epsilon_{p(he)}$, we perform calculations for several of their  randomly chosen combinations. We choose the escape parameter in the range of $\alpha=(1-3)$. This range approximately covers the $\alpha$ values given in some available literatures. Studies based on non-linear diffusive shock acceleration theories which takes into account the modification of the shock structure by the CRs give $\alpha\sim 0.8$ (e.g., Ptuskin $\&$ Zirakashvili 2005). Blasi $\&$ Amato 2011 adopted $\alpha\sim 3.2$ in their study of the effect of random nature of SNRs on the CR spectrum. Investigations of $\gamma$-ray emissions from molecular clouds interacting with nearby SNRs adopt values in the range of $\alpha=(2.4-2.6)$ (Gabici et al. 2009, Ohira et al. 2011). We consider the CR injection efficiencies in the range of $\epsilon_p=(5-25)\%$ for protons and $\epsilon_{he}=(1-5)\%$ for helium, where the values are in units of $10^{51}$ ergs. The averaged proton to helium injection ratio of 5 which we consider here is less than the observed proton to helium flux ratio of $\sim (20-13)$ in the energy range of $\sim (20-200)$ GeV/n (Yoon et al. 2011). But, our wide range of efficiencies for both the species well cover the observed flux ratios. It should be understood that the observed flux ratios may not necessarily represent the injection efficiency ratios from the source. Effects during the propagation in the Galaxy such as due to spallation (which are different for different nuclear species depending on their interaction cross-sections) may change the composition ratios produced by the source. In addition, propagation of CRs is charged dependent. Those which undergo faster diffusion will escape more easily from the Galaxy and eventually lead to less flux observed at the Earth.

Fig. 1 shows our calculated proton spectra $\left(\times E^{2.75}\right)$ for Models A (top panel) and B (bottom panel). In each panel, the thin black lines represent an example of $30$ different random spectra we have calculated. Each random spectrum corresponds to a set of $\{\alpha,\epsilon_{p(he)}\}$ which is assumed to be the same for all the SNRs. Each spectrum is the sum of the background CRs (shown as the blue dotted line) and the total contribution from all the nearby SNRs we have considered. The blue solid line represents the averaged spectrum of a total of $200$ such random spectra. The data are taken from CREAM (Yoon et al. 2011), ATIC\footnote{Data taken from the database compiled by Andrew W. Strong (Strong $\&$ Moskalenko 2009)} (Panov et al. 2007), AMS (Alcaraz et al. 2000, Aguilar et al. 2002), BESS (Haino et al. 2004), and PAMELA\footnotemark[\value{footnote}] (Adriani et al. 2011) experiments. One common result that we can notice between the two models is that the contribution of the nearby SNRs show up mostly above $\sim (0.5-1)$ TeV. However, there are some general differences between the two results. The results for Model A not only show larger variations between individual spectra but also stronger irregular features and spikes. Also, in general Model A produces larger contribution from the nearby SNRs as compared to Model B. This is largely due to the comparatively harder source spectrum of CRs required in Model A. For the reasonable range of injection efficiencies considered in our study, the results of Model A seem to be in better agreement with the data both in terms of the size and the shape of the spectra. On comparing the averaged spectra (thick blue lines) above $\sim (0.5-1)$ TeV, the result of  Model A is comparatively harder up to $\sim 100$ TeV which then becomes steeper at higher energies. This spectral behavior of Model A is in good agreement with the recent data which also seem to indicate that the spectral hardening for protons does not persist beyond $\sim (20-30)$ TeV. On the other hand, the averaged spectrum in Model B show less hardening above $\sim 1$ TeV and it continues without any turn over or steepening up to the maximum energy considered here.

The corresponding results for helium are shown in Fig. 2: Model A (top panel) and Model B (bottom panel). Our results for helium look similar to those obtained for protons. One general difference we notice is the shifting of the helium results towards lower energies with respect to the proton results. Though not very significant, a similar trend is also present in the observed data. For instance, the spectral hardening in the helium data occurs at $\sim 0.5$ TeV/n whereas for the protons it occurs at $\sim 1$ TeV. Moreover, the spectral turnover at higher energies seems to occur at $\sim 10$ TeV/n for helium while for protons it seems to occur at $\sim (20-30)$ TeV.

In Fig. 3, we present our best fit results: protons ($\times E^{2.75}$, top panel) and helium ($\times E^{2.7}$, bottom panel). They are obtained by choosing $(\alpha, \epsilon_p, \epsilon_{he})=(2.2, 9\%, 2\%)$ for Model A and $(2.4, 20\%, 3.7\%)$ for Model B. Our model parameters give escape times of $t_{esc}=(500-10^5)$ yr for protons of energies ($1$ PeV$-8.6$ GeV) and for helium of ($0.5$ PeV/n$-4.3$ GeV/n) in Model A. In model B, the corresponding values are ($1$ PeV$-3$ GeV) and ($0.5$ PeV/n$-1.5$ GeV/n) respectively. The data in Fig. 3 are the same as in Figs. 1 $\&$ 2 respectively. The blue dotted line represents the background CR spectrum. The solid lines correspond to Model A and the double dotted lines to Model B in which the thin and the thick lines represent the total contributions from the nearby local SNRs and the total background plus nearby contributions respectively. In Model A, the dominant nearby contributors are the Vela, G299.2-2.9 and SN185 remnants. They are shown by the thin dashed lines in the figures. Vela dominates in the range of $\sim (0.7-10)$ TeV/n while above that, the spectrum is mostly dominated by G299.2-2.9 and SN185. In Model B, Vela dominates over a wide range up to $\sim 300$ TeV/n and beyond that, it is dominated by G299.2-2.9 (not shown in the figure).

\begin{figure}
\centering
\includegraphics*[width=0.315\textwidth,angle=-90,clip]{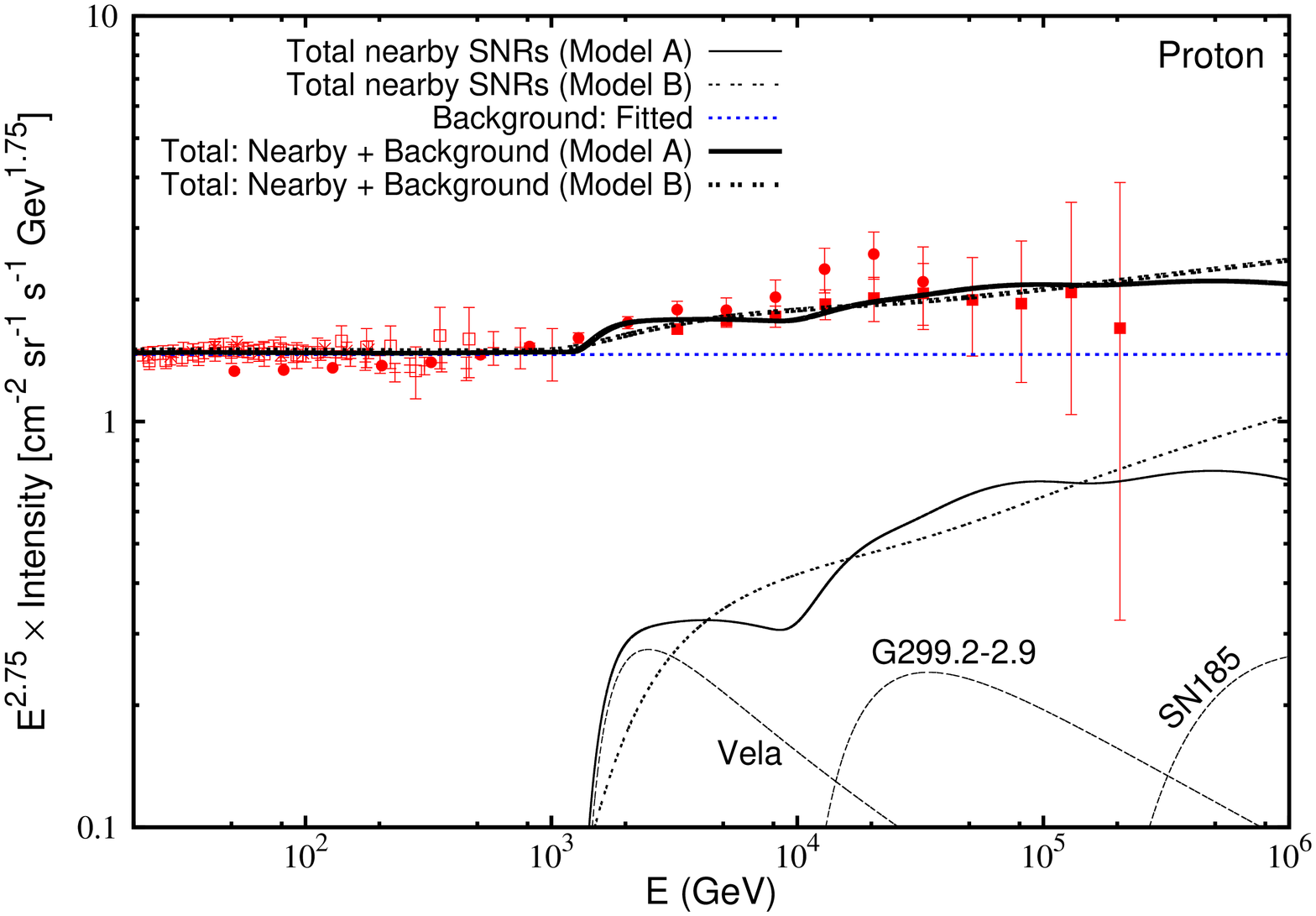}\\
\includegraphics*[width=0.315\textwidth,angle=-90,clip]{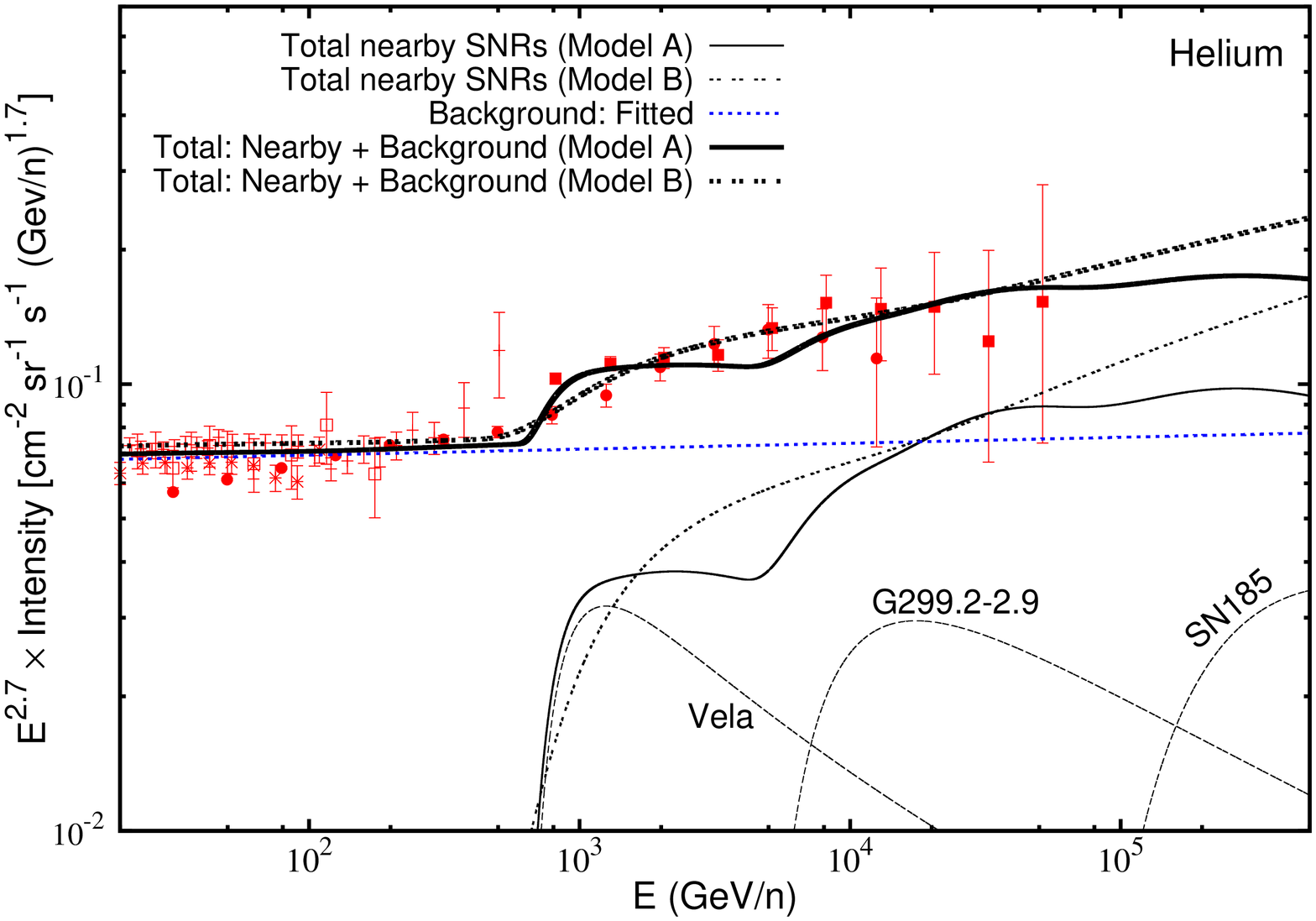}
\caption{\label {fig1} Best fit results for proton ($\times E^{2.75}$, top panel) and helium ($\times E^{2.7}$, bottom panel). The data are the same as given in Figs. $1\&2$ respectively. The blue dotted line represents the background spectrum. Solid lines correspond to Model A and double dotted lines to Model B of which the thin line represents the total contributions from the nearby SNRs and the thick line represents the total background plus nearby contributions. The thin dashed lines represent the dominant nearby contributors in Model A.}
\end{figure}

The steep low energy cut-offs in the individual SNR contributions in our model are largely due to the energy dependent escape of CRs. CRs below the cut-offs are mostly those which are still confined within the SNRs and are not yet released into the interstellar medium. Our best fit result for Model A show a rise in the total spectrum at $\sim (0.5-1)$ TeV/n which remain constant up to $\sim (5-10)$ TeV/n. This is due to the effect of the low energy cut-off of the Vela remnant. To be specific, the rise in the proton spectrum occurs at $\sim 1.2$ TeV while that of the helium at $\sim 0.6$ TeV/n. This difference is largely due to the effect of early escape of helium compared to the protons for the same energy/nucleon. There is also some effect due to the faster diffusion of helium than the protons for the same energy/nucleon. As mentioned in section 3.1, it is interesting to see that the recent data also seem to indicate that the helium spectrum starts hardening at comparatively lower energies than the protons. Our best fit results then show a slow increase above $\sim (5-10)$ TeV/n which again becomes almost constant above $\sim (40-60)$ TeV/n. This is due to the combined effect of other SNRs mainly G299.2-2.9 and SN185. These spectral features are found to be more pronounced for helium. In Model B, these features are smeared out and we get a smooth spectrum with a slow increase above a few TeVs. This is due to the comparatively slower diffusion of CRs in this model and the dominance by a single source (Vela) over a wide range of energy spectrum.

The total CR anisotropy $\Delta$ expected under our model can be calculated using the following equation (Thoudam 2007),
\begin{equation}
\Delta=\frac{\displaystyle\sum_{i} I_i\delta_i \hat{r}_i.\hat{r}_m}{I_T}
\end{equation} 
where the summation is over the nearby SNRs, $\hat{r}_i$ denotes the direction of the $i^{th}$ SNR giving an intensity $I_i$ at the Earth, $\hat{r}_m$ denotes the direction of maximum intensity, $I_T$ represents the total observed CR intensity and $\delta_i$ denotes the anisotropy amplitude due to a single SNR. $\delta_i$ under the diffusion approximation is given by (Mao $\&$ Shen 1972),
\begin{equation}
\delta_i=\frac{3D}{c}\frac{|\nabla N_i|}{N_i}
\end{equation}
where $N_i$ (given by Eq. 7) denotes the CR density from an SNR with distance $r_i$ and age $t_i$. For our best fit proton results, we get $\Delta\approx (1.7\times 10^{-2}-0.12)$ and $(1-4)\times10^{-2}$ for Model A and B respectively in the energy range of $(1-100)$ TeV. Our estimates are larger than the measured anisotropies of $\sim (0.5-1)\times 10^{-3}$ in the same energy range. But, compared to Model A, Model B looks more closer to the measured values (see also Ptuskin et al. 2006).

\section{Discussions and conclusions}
We show that for both Models A and B, the nearby SNRs contribute mostly above $\sim(0.5-1)$ TeV/n and they may account for the observed spectral hardening at high energies. We show this for a wide range of CR injection efficiencies  and CR escape parameters from the SNRs. Looking into the averaged spectra in Figs. $1\&2$, we find that both the models predict that the hardening of the helium spectrum should occur at lower energies as compared to the protons. We also find that the averaged result of Model A seems to explain the overall data better than that of Model B.

However, the wide range of parameters values considered in our study allow both the propagation models to successfully explain the observed data with a careful choice of model parameters. We show this with our best fit results in Fig. 3. But, the high CR injection efficiency of $\epsilon_p=20\%$ required in Model B is around a factor of $2$ larger than the normally considered value of $\sim 10\%$ for CR studies in the Galaxy. Moreover, the steep source index of $\gamma\sim 2.4$ required in this model is also hard to reconcile with the results of diffusive shock acceleration theory which predict an index of $\gamma= 2$. Model A, on the other hand, looks favorable considering its relatively more reasonable values of the source index $(\gamma= 2.15)$ and the proton injection efficiency $(\epsilon_p=9\%)$ required to explain the observed hardening. In addition, Model A also better explain the apparent observed property that the spectral hardening does not persist above a few TeVs. However, the measured anisotropy seem to favor Model B which assumes a weaker energy dependence of CR diffusion in the Galaxy.

Our results look different from the predictions of other models. Models based on constant diffusion coefficient at high energies or spectral dispersion in the source spectrum are expected to produce a high energy spectrum which remains hard up to the maximum energy (Ave et al. 2009). But, the data indicates that the spectral hardening happens only up to $\sim (20-30)$ TeV for protons and $\sim 10$ TeV/n for helium which in general agrees well with our predictions (especially with Model A). It should be mentioned that our results may not be significantly different from others if the CR spectrum has a break or a cut-off (normally assumed to be exponential) at energies $\lesssim 0.1$ PeV. In such a case, the spectrum will start rolling over before it starts showing noticeable differences. But, note that a cut-off somewhere between $\sim (3-5)$ PeV is preferred, irrespective of the nature and the origin of the cut-off, in order to explain the observed knee in the energy spectrum of CRs (H\"orandel 2003).

The secondary CR spectrum under our model can be even more different from other models. Secondaries are those which are considered to be produced by the spallation of the primaries only during the propagation in the Galaxy. Their spectrum $N_s$ in the Galaxy is related to their primary spectrum $N_p$ as $N_s(E)\propto N_p(E)/D(E)$. Thus, for $N_p(E)\propto E^{-\Gamma}$, the secondary spectrum follows $N_s(E)\propto E^{-(\Gamma+a)}$ which is steeper than their primaries by the diffusion index $a$. Therefore, once $D(E)$ is fixed, the shape of the secondary spectrum depends on the shape of their primary spectrum. This means that models which assume the same $D(E)$ but different $N_p(E)$ will produce different $N_s(E)$. Under our model, if we neglect the production of secondaries inside the SNRs, we can assume that all the secondaries are produced by the background CRs. As our background primary spectrum is steeper than the spectrum used in other models to explain the spectral hardening, e.g. Yuan et al. 2011, we expect a steeper secondary spectrum in our case. This difference would be even more significant when compared to propagation models which assume a constant CR escape time from the Galaxy at higher energies (Ave et al. 2009). Under such models, $N_s(E)\propto E^{-\Gamma}$ at higher energies while at lower energies $N_s(E)\propto E^{-(\Gamma+ a)}$. The differences we just mentioned are expected in all kinds of secondary nuclear species like boron, sub-Fe, and anti-protons. At present, data on secondary spectra are available at most only up to $\sim 100$ GeV/n. Future high energy measurements would be crucial to test our model.

In addition, the diffuse $\gamma$-ray emission of our Galaxy can also provide an important check of our model. If the diffuse emission is dominated by the $\pi^0$-decay $\gamma$-rays, then their intensity would largely follow the proton spectrum at high energies. Therefore, under our model we expect a diffuse spectrum which is steeper than the predictions from other models. In fact, it has already been shown in Yuan et al. 2011 that under their model, the $\gamma$-ray spectrum would become harder above $\sim 50$ GeV. Preliminary results from the FERMI measurements up to $\sim 100$ GeV show that the spectrum is in good agreement with the conventional model assuming a single power-law CR spectrum above a few GeV (Strong 2011). The spectrum do show some excess above the model which could well be attributed to unresolved point sources like pulsars. Detailed investigation of the diffuse $\gamma$-ray spectrum and also future measurements at even higher energies would be important to check the validity of our model.

In short, we conclude that the apparent change in the spectral index of the CR energy spectra at TeV energies could be a local effect due to nearby SNRs. A detailed investigation of both the proton and the helium spectra seems to favor this model. Future measurements of secondary CR spectra and of the Galactic diffuse $\gamma$-ray emission at TeV energies can provide a deeper understanding of the problem.
\\
\\
\textbf{REFERENCES}\\
\\
Adriani, O., et al. 2011, Science, 332, 69\\
Aguilar, M., et al. 2002, Physics Reports 366, 331\\
Aharonian,F. A., et al. 2006, ApJ, 636, 777\\
Aharonian, F. A., et al. 2008a, A$\&$A, 477, 353\\
Aharonian, F. A., et al. 2008b, PRL, 101, 261104\\
Ahn, H. S. et al. 2010, ApJL, 714, L89\\
Alcaraz, J., et al. 2000, Physics Letters B 490, 27\\
Ave, M. et al. 2008, ApJ, 678, 262\\
Ave, M. et al., 2009, ApJ, 697, 106\\
Bell, A. R., 1978, MNRAS 182, 147\\
Berezhko, E. G. 1996, APh, 5, 367\\
Biermann, P. L., Becker, J. K., Dreyer, J., Meli, A., Seo, E., $\&$ Stanev, T. 2010, ApJ, 725, 184\\
Blandford, R., $\&$ Eichler, D., 1987, Physics Reports, 154, 1\\
Blasi, P., $\&$ Amato, E., arXiv:1105.4521\\
Caprioli, D., Amato, E. $\&$ Blasi, P., 2010, APh, 33, 160\\
Caprioli, D., Blasi, P., $\&$ Amato, E. 2009, MNRAS, 396, 2065\\
Erlykin, A. D. $\&$ Wolfendale, A. W., 2011, $32^\mathrm{nd}$ ICRC, Beijing\\
Gabici, S., Aharonian, F. A., $\&$ Casanova, S. 2009, MNRAS, 396, 1629\\
Haino S. et al., 2004, Phys. Lett. B, 594, 35\\
H\"orandel, J. R., 2003, APh, 19, 193\\
Mao, C. Y., $\&$ Shen, C. S., 1972, Chin. J. Phys., 10, 16\\
Ohira, Y., Murase, K., $\&$ Yamazaki, R., 2010, A$\&$A, 513, A17\\
Ohira, Y., $\&$ Ioka, K., 2011, ApJ, 729, L13\\
Ohira, Y., Murase, K., $\&$ Yamazaki, R. 2011, MNRAS, 410, 1577\\
Panov, A. D. et al. 2007, Bull. Russ. Acad. Sci., Vol. 71, No. 4, pp. 494\\
Parizot, E., Marcowith, A., Ballet, J., $\&$ Gallant, Y. A. 2006, A$\&$A, 453, 387\\
Ptuskin, V. S., Jones, F. C., Seo, E. S., $\&$ Sina, R., 2006, AdSpR, 37, 1909\\
Ptuskin, V. S., $\&$ Zirakashvili, V. N., 2005, A$\&$A, 429, 755\\
Strong, A. W., 2011, arXiv:1101.1381\\
Strong, A. W. $\&$ Moskalenko, I. V., 2009, $31^\mathrm{th}$ ICRC, LODZ\\
Strong, A. W., et al. 2010, ApJL, 722, L58\\
Thoudam, S. 2007, MNRAS, 378, 48\\
Thoudam, S. 2008, MNRAS, 388, 335\\
Thoudam, S., $\&$ H\"orandel, J. R. 2011, arxiv:1109.5588\\
Yoon, Y. S. et al. 2011 ApJ 728 122\\
Yuan, Q., Zhang, B., $\&$Bi, X. -J, 2011, Phys. Rev. D 84, 043002\\
Vladimirov, A. E, J\'ohannesson, G., Moskalenko, I. V. $\&$ Porter, T. A., 2011, arxiv:1108.1023
\end{document}